\documentclass[12pt,prl,aps,nofootinbib]{revtex4}
\usepackage{epsfig,epsf}
\usepackage{bm}
\begin{document}                                                              
\newcommand{\beq}{\begin{equation}}
\newcommand{\eeq}{\end{equation}}
\newcommand{\exponential}[1]{\mathrm{e}^{#1}}
\newcommand{\be}{\begin{eqnarray}}
\newcommand{\ee}{\end{eqnarray}}
\newcommand{\dd}{\mathrm{d\,}}
\def\eq#1{{Eq.~(\ref{#1})}}
\def\fig#1{{Fig.~\ref{#1}}}
\newcommand{\as}{\alpha_S}
\newcommand{\bas}{\bar{\alpha}_S}
\newcommand{\h}{\frac{1}{2}}
\newcommand{\bra}[1]{\langle #1 |}
\newcommand{\ket}[1]{|#1\rangle}   
\newcommand{\bracket}[2]{\langle #1|#2\rangle}
\newcommand{\intp}[1]{\int \frac{d^4 #1}{(2\pi)^4}}
\newcommand{\mn}{{\mu\nu}}
\newcommand{\ab}{{\alpha\beta}}
\newcommand{\tr}{{\rm tr}}
\newcommand{\Tr}{{\rm Tr}}
\newcommand{\T} {\mbox{T}}
\newcommand{\braket}[2]{\langle #1|#2\rangle}
\newcommand{\sigin}{\sigma_{in}}

\begin{flushright}
BNL-NT-04/26\\
August 3, 2004
\end{flushright}

\title{Color Glass Condensate at the LHC: \\  hadron multiplicities in $pp$, $pA$ and $AA$ collisions}

\author{Dmitri Kharzeev$^a$, Eugene Levin$^{a,b}$ and Marzia Nardi$^{a,c}$}

\bigskip
 
\affiliation{
a) Department of Physics, Brookhaven National Laboratory,\\
Upton, New York 11973-5000, USA\\
b) HEP Department, School of Physics,\\
Raymond and Beverly Sackler Faculty of Exact Science,\\
Tel Aviv University, Tel Aviv 69978, Israel\\
c) Dipartimento di Fisica Teorica dell'Universit{\`a} di Torino,\\ 
via P.Giuria 1, 10125 Torino,
     Italy}


\begin{abstract}

We make quantitative predictions for the rapidity and centrality dependencies of 
hadron multiplicities in $AA$, $pA$  and $pp$ collisions at the LHC energies basing on the ideas of 
parton saturation in the Color Glass Condensate. 
 
\end{abstract}
\maketitle

\section{Introduction}

At high energies QCD is expected to enter the new phase :
the Color Glass Condensate (CGC) which is characterized by strong coherent gluon fields leading to parton saturation \cite{GLR,hdQCD,MV,BK,JIMWLK}. 
Previously, we have applied this approach  \cite{KN,KL,KLN,KLM,KKT} to 
describe the wealth 
of experimental data \cite{Brahms,Phenix,Phobos,Star} from RHIC. 
The LHC will allow to extend further the investigations of QCD in the regime of high parton density. This is because the new scale of the problem, the saturation 
momentum $Q_s$,  will become so large ($Q^2_s \,\approx\,5 - 10\ {\rm GeV}^2$) that a separation of CGC physics from non-perturbative effects 
should become easier. 
The main objective of this paper is to give predictions for the global characteristics of the inelastic events  
in nucleus-nucleus, proton-nucleus and proton-proton collisions at LHC energies basing on the ideas of parton saturation in the 
Color Glass Condensate (CGC). 

To understand better the differences implied by a higher energy of the LHC, let us start with the main 
assumptions of the approach we used to describe the data from RHIC:
\begin{enumerate}
\item \quad 
At Bjorken $x\, \leq\, 10^{-2}\,$ the inclusive production of partons (gluons and quarks) is driven by parton saturation in 
strong gluon fields as given by
McLerran-Venugopalan model \cite{MV}. 

\item \quad 
The region of $x \,\approx\,10^{-3}$ (accessible at forward rapidities at RHIC) is considered as the low $x$ region in which
$\as\,\ln(1/x) \,\approx\,1$ so the quantum evolution becomes important; we assume that     $\as \,\ll\,1$ to keep the calculation 
simple and transparent;

\item \quad 
We assume that the interaction in the final state does not change significantly 
the  multiplicities of partons resulting from the early stages of the process; 
this may be a consequence of local parton hadron duality,  
or of the entropy conservation. Therefore multiplicity measurements are extremely 
important for uncovering the reaction dynamics.  However, we would like to state clearly that we do not claim 
that the interactions in the final state are unimportant. Rather, we consider the CGC as the initial condition for the 
subsequent evolution of the system, which can be described for example by means of 
hydrodynamics (such an approach has been followed in Refs. \cite{Eskola,Hirano}).  

\end{enumerate}

Even a superficial glance at these three assumptions reveals that the conditions for the applicability of our approach at the LHC 
improve.
Indeed, at LHC energies the value of $x$ will be two orders of magnitude lower 
than at RHIC. This makes the use of the well--developed methods of low $x$ physics \cite{MV1,BK,JIMWLK,LL,IM,KL,MUSO} better justified. At  LHC energies we have a  theoretical tool to deal with the high parton density QCD   
in the mean field approach ( so called Balitsky-Kovchegov non-linear equation \cite{BK}), or on a general basis 
of the JIMWLK equation  \cite{JIMWLK};  even more general approaches may be possible (see for example the Iancu-Mueller factorization 
\cite{IM,KLE}). However, despite a number of well developed approaches which could be applied at low $x$ we would 
like to warn that even the LHC energy is not high enough to apply any of the methods mentioned above  
without discussing possible "pre-asymptotic" corrections to them.
 
Consider for example the determination of the value of the saturation momentum -- the 
key scale in the CGC phase of QCD. As was noticed first in Ref. \cite{DIO} the value of the saturation scale is 
affected by the next-to-leading order corrections to the BFKL kernel which were neglected in all of the discussed above 
approaches. Their numerical significance is so large that they cannot be neglected: if the next-to-leading order BFKL kernel is used, 
because of a large energy extrapolation interval to the LHC the value of 
$Q^2_s$   turns out to be 5 - 10 times smaller than if one uses the 
leading order kernel ( see detailed  discussion in Ref. \cite{DUR}).  However the good news is that the NLO corrections appear under 
theoretical control and we can take them into account. 

\vskip0.3cm

The paper is organized as follows. In the second section we discuss the geometry of nucleus-nucleus and hadron-nucleus collisions and introduce 
the Glauber formalism we use.  In the third section, we review the general formalism which we use to evaluate the multiplicities; we also discuss 
the influence of higher order corrections and the effects of the running coupling constant on the results. In the fourth section we 
list the parameters of our approach and justify the values we use; we then give a complete set of predictions for hadron multiplicities 
at the LHC energies in $Pb-Pb$, $p-Pb$, and $pp$ collisions, including the dependences on rapidity and centrality.  We then summarize our 
results.

\section{The geometry of nucleus-nucleus and hadron-nucleus collisions and the Glauber approach}

At high energies the paths of the colliding nucleons can be approximated by straight lines, since in a typical interaction $t/s \ll 1$ and 
the typical scattering angle is small. This is the most important approximation underlying the Glauber approach to nuclear interactions. 
Other approximations which simplify calculations but are in principle unnecessary are the smallness of the nucleon--nucleon interaction 
radius compared to the typical nuclear size, and the neglect of the real part of the $NN$ scattering amplitude.   
Many quantities characterizing the geometry of the collision can be readily computed in this approach; a complete set of the relevant formulae can be found e.g. in \cite{KNS} and we will not reproduce all of them here.

It is customary and convenient to parameterize the centrality of the collision in terms of the "number of participants" $N_{part}$ -- the number 
of nucleons which underwent at least one inelastic collision. This number can be directly measured experimentally (at least in principle) 
by detecting  in the forward rapidity region the number of "spectator" nucleons $N_{spect}$ which did not take part in any inelastic collisions; 
obviously, for a nucleus with mass number $A$,  $N_{part} = A - N_{spect}$.  

The number of participating nucleons in a 
nucleus-A--nucleus-B interaction
depends on the impact parameter $b$. In the eikonal approximation it
can be evaluated as (see \cite{WNM}):

\be
N_{part}^{AB}(b) = \int d^2 s \, n_{part}^{AB}({\bf b},{\bf s})
 &=& A\, \int d^2 s \, T_A({\bf s})
\left\{1-\left[1-\sigin \, T_B({\bf b} -{\bf s})\right]^B \right\} 
\nonumber \\
&& + \, B\,  \int d^2 s\, T_B( {\bf b} -{\bf s})
\left\{1-\left[1-\sigin \, T_A({\bf s})\right]^A \right\} \,  ,
\label{npartAB}
\ee
with the usual definition for
the nuclear thickness function $T_A({\bf s})= \int_{-\infty}^\infty d z
\rho_A(z,{\bf s})$, normalized as $\int d^2s\, T_A({\bf s}) = 1$; 
$\sigin$ is the proton-proton inelastic cross-section 
without diffractive component. For the LHC energies we assumed
$\sigin = 70$~mb (\cite{GLMSF}).

From \eq{npartAB} the definition of the local density of participants
$n_{part}^{AB}({\bf b},{\bf s})$ is evident; 
we will define its 
average over the transverse plane as 
\beq
\langle n_{part}^{AB}\rangle ( b ) =
\frac{ \int d^2 s \, [n_{part}^{AB}({\bf b},{\bf s})]^2}
     { \int d^2 s \, n_{part}^{AB}({\bf b},{\bf s}) }~.
\label{densAB}
\eeq
In the following we will need to use the average number of
participants computed separately for nucleus-A and nucleus-B; 
it is given by
\beq
\langle n_{part,A}^{AB}\rangle ( b ) =
\frac{ \int d^2 s \, n_{part,A}^{AB}({\bf b},{\bf s}) \,
                            n_{part}^{AB}({\bf b},{\bf s})}
     { \int d^2 s \, n_{part}^{AB}({\bf b},{\bf s}) }~.
\eeq
Obviously, one has for their sum  
\[
\langle n_{part,A}^{AB}\rangle ( b ) +
\langle n_{part,B}^{AB}\rangle ( b ) = \langle n_{part}^{AB}\rangle ( b )
\]
where $\langle n_{part,A}^{AB}\rangle ( b )$ 
and $\langle n_{part,B}^{AB}\rangle ( b )$ 
are the integrands of the
first term and  second term in the r.h.s of \eq{npartAB} respectively.

In table \ref{tab_b} we give the number of participants and 
their density (respectively Eqs. (\ref{npartAB}) and (\ref{densAB}))
for Pb-Pb collisions at LHC. 

\begin{center}
\begin{table}
\begin{tabular}{c|c|c||c|c|c}
$ \begin{array}{c} b \\ ({\mathrm{fm}}) \end{array} $
& $N_{part}^{AB}$ &
$ \begin{array}{c} n_{part}^{AB} \\ ({\mathrm{fm}}^{-2}) \end{array}$ &
$ \begin{array}{c} b \\ ({\mathrm{fm}}) \end{array} $
& $N_{part}^{AB}$ &
$ \begin{array}{c} n_{part}^{AB} \\ ({\mathrm{fm}}^{-2}) \end{array}$\\
\hline
 0.00 &  406.9  &     2.98 &   8.00 &  166.8  &     2.21 \\
 1.00 &  402.4  &     2.97 &   9.00 &  127.5  &     1.97 \\
 2.00 &  387.8  &     2.93 &  10.00 &  91.9   &     1.69 \\
 3.00 &  363.2  &     2.88 &  11.00 &  61.1   &     1.35 \\
 4.00 &  330.3  &     2.80 &  12.00 &  36.2   &     0.98 \\
 5.00 &  291.9  &     2.70 &  13.00 &  18.3   &     0.59 \\
 6.00 &  250.6  &     2.57 &  14.00 &  7.5    &     0.27 \\
 7.00 &  208.3  &     2.41 &  15.00 &  2.5    &     0.09 \\
\hline
\end{tabular}
\caption{Mean number of participants and their average density in Pb-Pb 
collisions at LHC as a function of $b$}
\label{tab_b}
\end{table}
\end{center}

The corresponding formulae for the proton--nucleus $pA$ interaction can be 
deduced by setting $B=1$ and using a delta-function for the 
proton thickness function (in the point-like approximation for the size of the proton).
We get from \eq{npartAB}:
\be
N_{part}^{pA}(b) = 
 A\, \sigin \, T_A( b) +
\left\{1- \left[ 1-\sigin \, T_A( b) \right]^A \right\} \, =
A\, \sigin \, T_A( b) + \left\{1-P_0^{pA}(b) \right\}.
\label{npartpA}
\ee

In the previous formula the function $P_0^{pA}(b)$ is the 
probability of no interaction in a p-A collision at impact parameter $b$;
the integration of $[1-P_0^{pA}(b)]$ over $b$ gives the inelastic
proton-nucleus cross section $\sigma_{pA}$.

The average number of participants in a p-A collision can be obtained as:
\beq
\langle N_{part}^{pA} \rangle = \frac{\int \dd^2 b \, N_{part}^{pA}(b) }
{\int \dd^2 b  \, [1-P_0(b)]} =
A\, \frac{\sigma_{in}}{\sigma_{pA}} + 1;
\label{meanNppA}
\eeq
 the first term in the r.h.s. gives the mean number of participants 
$\langle N_{part,A}^{pA}\rangle $ in 
the nucleus. As in the case of nucleus-nucleus collision, we will need
to compute the density of participants in nucleus A, defined as:
\beq
\langle n_{part,A}^{pA} \rangle = \frac{\langle N_{part,A}^{pA}\rangle}
{\sigin}= \frac{A}{\sigma_{pA}}.
\label{denspA}
\eeq

In practice, the information about
the impact parameter dependence is extracted by analyzing  the data 
in various centrality bins. The physical observable most frequently used 
to estimate the centrality of the collision is the multiplicity of 
charged particles $N_{ch}$. 
We will assume that the average value of $N_{ch}$ produced in a collision
at impact parameter $b$ is determined by the number of participating 
nucleons $N_{part}(b)$. The actual multiplicity will 
fluctuate around its mean value according to:
\beq
{\mathcal P}(N_{ch},\langle N_{ch}(b)\rangle )=
\frac{1}{\sqrt{2 \pi a \langle N_{ch}(b)\rangle}} \,
C(\langle N_{ch}(b)\rangle) \,
\mathrm{exp} \left\{ - \frac{[N_{ch}-\langle N_{ch}(b)\rangle]^2}
{2a\langle N_{ch}(b)\rangle }\right\},
\eeq
where the factor $C(N) \equiv 2/[1+erf(\sqrt{N/2a})]$ is
introduced to ensure that the fluctuation function 
${\mathcal P}(N_{ch}, N )$ satisfies $\int_0^\infty dN_{ch}
{\mathcal P}(N_{ch}, N ) = 1$. The numerical value of $C(N)$
is 1 with very good accuracy for almost all cases of practical 
interest (it can exceed 1 for very peripheral collisions,
where the number of participants and consequently $N_{ch}$ is small:
in such a case it is important to include the factor $C(N)$ to have a correct
normalization).

The parameter $a$ gives the width of the fluctuations: its value is dependent 
on the experimental apparatus, therefore it is not
possible for us to predict its value for the LHC experiments. For the experiments 
at SPS and RHIC the value of $a$ varies from 0.5 to 1.5-2.
We will assume 
$a=0.5$ in the following; uncertainty in this parameter can affect the centrality dependence of our results. 
In the case of $Pb Pb$ collisions, we estimate the resulting uncertainty in the density of participants (and thus 
in the saturation scale, see below) to be about $5 \%$; in the case of $p Pb$ collisions, this uncertainty can reach 
$10 \div 15 \%$ for peripheral collisions.

\medskip

We will also assume the proportionality between $N_{ch}$ and $N_{part}$ when computing 
the differential inelastic cross section; this proportionality is not exact, but
the shape of minimum bias distribution of events
which is normally used to fix the parameter $a$
(and the proportionality constant between $N_{ch}$ and $N_{part}$)
 has been found insensitive to this assumption (see \cite{KN}).

The minimum bias differential cross section can be obtained as ($N(b)\equiv q N_{part}(b)$, where $q$ is a constant):
\beq
\frac{d\sigma_{mb}}{dN_{ch}}=\int d^2b \,
{\mathcal P}(N_{ch}, N(b) ) \, \left[1-P_0(b) \right];
\label{minbias}
\eeq
here $P_0(b) $ is the probability of no interaction at the impact parameter $b$:
for a nucleus-nucleus collision $P_0(b)=[1-\sigin T_{AB}(b)]^{AB}$ 
where $T_{AB}$ is the overlap function : 
$T_{AB}(b)=\int d^2s \, T_A(s)T_B({\bf b} -{\bf s})$; in the case of 
B=1, $P_0(b)$ reduces to $P_0^{pA}(b)$ defined above.
In the following, all of the formulae will refer to A-B collisions;
with obvious modifications they are valid also in the p-A case.

The total nucleus-nucleus cross section is then obtained by 
integrating \eq{minbias} over $dN_{ch}$: 
\beq
\sigma_{AB} = \int dN_{ch} \, \frac{d\sigma_{mb}}{dN_{ch}}=
\int d^2b  \, \left[1-P_0(b) \right].
\eeq

The mean value of any physical observable ${\mathcal O}$ 
(given in terms of the impact parameter $b$) can be computed as:
\beq
\langle {\mathcal O}\rangle = \frac{1}{\sigma_{AB} }
 \int dN_{ch}  \, \frac{d\sigma_{mb}}{dN_{ch}} \,
{\mathcal O}(b) 
\eeq

To obtain the corresponding average for a given centrality cut
we have to limit the integrations in the previous formula in the
appropriate way, for instance the expression:
\be
\left. \langle {\mathcal O}\rangle \vphantom{\frac{1}{1}}
\right|_{N_{ch}>N_0}
= \frac{\displaystyle
 \int_{N_0} dN_{ch}  \, \frac{d\sigma_{mb}}{dN_{ch}} \, {\mathcal O}(b) }
{\displaystyle \int_{N_0} dN_{ch}  \, \frac{d\sigma_{mb}}{dN_{ch}} },
\ee
gives the average value of the observable ${\mathcal O}$ in the
fraction of the total cross section defined by the limit $N_0$.
In this work the previous formula has been used to compute the mean density
of participating nucleons (\eq{densAB}) in different centrality bins,
as shown in table \ref{tab_cc}.

\begin{center}
\begin{table}
\begin{tabular}{r@-l@{\%~}|c|c}
\multicolumn{2}{c|}{centr. cut} & $\langle N_{part}^{AB} \rangle $ 
& $ \begin{array}{c}
\langle n_{part,A}^{pA} \rangle \\ ({\mathrm{fm}}^{-2}) \end{array} $
\\
\hline
  0 & 100 & 103.2 & 1.33 \\
\hline
  0 &   6 & 369.0 & 2.89 \\
  0 &  10 & 346.6 & 2.83 \\
  0 &  25 & 274.2 & 2.62 \\
 25 &  50 & 103.7 & 1.75 \\
 50 &  75 &  27.0 & 0.76 \\
 75 & 100 &   3.9 & 0.14 \\
\hline
  0 &  50 & 186.7 & 2.17 \\
 50 & 100 &  15.7 & 0.45 \\
\hline
\end{tabular}
\caption{Mean number of participants and their density in Pb-Pb 
collisions at LHC for different centrality bins}
\label{tab_cc}
\end{table}
\end{center}

Table \ref{tab_pAcc} gives the results of \eq{meanNppA} for the case of 
p-Pb collisions at LHC energy. The corresponding densities are obtained
according to \eq{denspA}.

\begin{center}
\begin{table}
\begin{tabular}{r@-l@{\%~}|c}
\multicolumn{2}{c|}{centr. cut} & $\langle N_{part}^{AB} \rangle $\\
\hline
  0 & 100 & 7.41 \\
\hline
  0 &  20 & 13.07 \\
  0 &  50 & 11.31 \\
 20 &  50 & 10.29 \\
 50 & 100 & 3.58 \\
\hline
\end{tabular}
\caption{Mean number of participants in p-Pb 
collisions at LHC for different centrality bins}
\label{tab_pAcc}
\end{table}
\end{center}


\section{The general formulae}

Let us discuss the main features of the approach we use to describe the production dynamics. As in our previous papers \cite{KL,KLN,KLM,KLN}
we use
 the following formula for the inclusive production \cite{GLR,GM}: 
\beq \label{INCR}
E {d \sigma \over d^3 p} = {4 \pi N_c \over N_c^2 - 1}\ {1 \over p_t^2}\  \times
\, \int^{p_t} \, d k_t^2 \,
\alpha_s \ \varphi_{A_1}(x_1, k_t^2)\ \varphi_{A_2}(x_2, (p-k)_t^2), \label{gencross}
\eeq
where $x_{1,2} = (p_t/\sqrt{s}) \exp(\mp y)$ and
$\varphi_{A_1,A_2} (x, k_t^2)$ is the unintegrated gluon distribution of a nucleus ( for the case of the proton 
one 
of $ \varphi_{A}$ should be replace by $ \varphi_{p}$.)
This distribution is related to the gluon density by 
\be \label{XGPHI}
xG(x,Q^2) \,\,=\,\,\int^{Q^2}\,\,d \,k^2_t \, \varphi(x, k_t^2).
\ee
We can compute the multiplicity distribution by integrating \eq{gencross} over $p_t$,
namely,
\beq \label{MULTI}
\frac{d N}{d y}\,\,=\,\,\frac{1}{S}\,\int\,\,d^2 p_t E {d \sigma \over d^3 p};
\eeq
$S$ is either the inelastic cross section for the minimum bias multiplicity, or a fraction of it 
corresponding to a specific centrality cut.

\subsection{Saturation scale}

Let us 
define two saturation scales: one for the nucleus $A_1$ and another for the nucleus $A_2$. We will see below that even in the case of 
$A_1 = A_2$ the introduction of two saturation scales will be useful.
  It is convenient to introduce two auxiliary variables, namely

\be 
Q_{s, min} (y,W) \,\,&=&\,\, min \left(Q_s(A_1;x_1=\frac{p_t}{W}\,e^{-y}), Q_s(A_2;x_2 
=\frac{p_t}{W}\,e^{y})\right)\,\,; \nonumber \\
 Q_{s, max} (y,W) \,\,&=&\,\, max  \left(Q_s(A_1;x_1=\frac{p_t}{W}\,e^{-y}), 
Q_s(A_2;x_2=\frac{p_t}{W}\,e^{y})\right)\,\,. \label{SCLS}
\ee

To understand the physical meaning of these two scales we start with the explicit formula for $Q_s$ which was 
suggested in Ref. \cite{GW} for the description of HERA data on deep inelastic scattering and was successfully 
used to describe the data from RHIC \cite{KN,KL,KLM,KLN}:

\beq \label{QS}
Q^2_s(x) \,\,=\,\,\,Q^2_0\,\left(\,\frac{x_0}{x}\,\right)^{\lambda}
\eeq
with the central value of $\lambda \,\,=\,\,0.288$ \cite{GW}; the value of $\lambda$ has an uncertainty of $5-10 \%$. 
Substituting $x_1 \,\,=\,\,(Q_s/W)\,e^{-y}$ and $x_2 \,\,=\,\,(Q_s/W)\,e^{y}$,
where $W$
is the energy of interaction, one can see that the energy and rapidity dependence of the saturation scale can be
reduced to a simple formula
\be \label{QSWY}
Q^2_s(A,y,W)\,\,=\,\,Q_0^2(A;W_0)\,\left( \frac{W}{W_0}\,e^{ y} \right)^{\frac{\lambda}{1 + \h \lambda}} 
\,\equiv\,Q_0^2(A;W_0)\,\left( \frac{W}{W_0}\, \right)^{\tilde{ \lambda}} \,e^{ \tilde{\lambda}\,y}\,\,.
\ee
In what it follows we will use the notation $\lambda $ for $ \tilde{ \lambda}\,=\, \lambda/(1 + \h \lambda) \, =\, 0.252$ hoping that it will not 
lead to misunderstanding.

\begin{figure}
\begin{center}
\epsfxsize=14.5cm
\hbox{ \epsffile{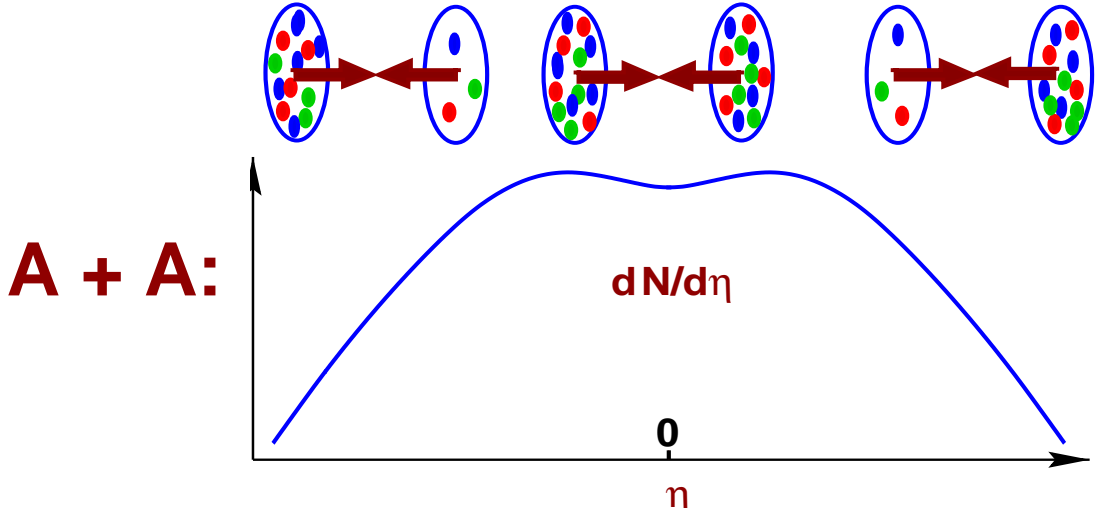}}
\end{center}
\caption{The CGC approach for nucleus - nucleus collision with  the saturation of parton density.}
 \label{aadist}
\end{figure}

Using \eq{QSWY} one can see that for a production of the gluon mini-jet at rapidities $y \neq 0$ there are two 
different saturation momenta: $Q^2_s(A;y,W)$ and $Q^2_s(A; -y,W)$, even for the collision of identical nuclei
(see \fig{aadist}). \fig{aadist} shows that the density is quite different in two nuclei since at $y \neq 0$ (say 
$y>0$)  one of 
the nuclei probed at relatively large $x = x_1 > x_2 $ is a rather dilute parton system while the second nucleus has much higher parton density than 
at $y =0$.  Therefore, for an $A + A$ collision  at $y > 0$    $Q_{s, min}  = Q_s(A;-y,W)$ while $Q_{s, 
max} = Q_s(A;y,W)$.  In the case of a collision of two different nuclei we need to take into account the 
$A-$dependent values of $Q_0(A;W)$  in \eq{QSWY}. 
 
The saturation scale is the main parameter of our approach and we need to understand clearly the energy 
dependence of $Q_s$ if we want to make predictions for the LHC energies.  The first basic result on the 
behavior of this scale is the power-like energy dependence  which follows directly from QCD for fixed QCD 
coupling.  As was shown in a number of papers \cite{GLR,BALE,BSB,BKL,MUTR,MP2003,DUR} the energy dependence of the 
saturation scale does not depend on the details of the behavior of the parton system in the saturation domain but 
can be determined just by using the perturbative QCD approach in the BFKL region \cite{BFKL}. Indeed, consider the 
 dipole-target scattering amplitude in the double Mellin transform representation, namely,
\be \label{MELLIN}
N(y,r^2)\,\,=\,\,\int \,\frac{d \omega \,d \gamma}{( 2 \pi i)^2}\,e^{ \omega\,\ln(1/x)\,\,+\,\,(\gamma - 
1)\, \ln(r^2 \Lambda^2_{QCD})}\,\,N(\omega, \gamma)\,\,.
\ee
The BFKL equation determines the value of $\omega$ at which $N(\omega, \gamma)$ has a pole: 
\be \label{BFKLOM}
\omega\,=\,\bas \,\chi(\gamma)
\ee
with a specific function $\chi$ which can be found e.g. in Ref. \cite{DUR}; we denote $\bas \equiv N_c \as/\pi$.  
To find the energy 
dependence of the saturation scale  we  first need to find a critical value of $\gamma=\gamma_{cr}$ defined by the 
equation \cite{GLR,MUTR,MP2003}
\be \label{SC1}
\frac{\chi(\gamma_{cr})}{1 - \gamma_{cr}}\,\,=\,\,- \,\frac{d \chi(\gamma_{cr})}{d \gamma}\,\,.
\ee
The meaning of this equation is the following: in the semi-classical approximation (see Ref. \cite{BKL} and references 
therein)
the scattering amplitude $N(y, \ln(r^2 \Lambda^2_{QCD}))$ has the following form:
\be
N(y, \xi \equiv \ln(r^2 \Lambda^2_{QCD})) = {\rm const}\times  \ \exp \left[\omega(y, \xi) y - (1-\gamma(y,\xi)) \xi \right].
\ee
The boundary of the saturation region is determined by the unique (critical) trajectory for the non-linear evolution equation in the 
$(y, \xi)$ plane for which the phase $v_{phase} = \omega(y, \xi) / (1 - \gamma(y, \xi))$ and the group 
$v_{group} = - d\omega(y, \xi) / d \gamma(y, \xi)$ velocities are equal. The physical meaning of this trajectory can be illustrated by an analogy 
in geometrical optics: the boundary which it defines is similar to the focal reflecting surface (therefore, one can see that the surface of the 
Color Glass shines!).   
The equality of phase and group velocities thus gives the equation for the saturation scale: 
\be \label{SC2}
\frac{ d \ln( Q^2_s(x)/\Lambda^2_{QCD})}{d \,\ln(1/x)}\,\,=\,\,\bas\, \frac{\chi(\gamma_{cr})}{1 - \gamma_{cr}} \, \equiv \, \lambda
\ee

For fixed $\as$ \eq{SC2} leads to 
\beq \label{SC3}
Q^2_s(x)\,\,=Q^2_0 \,\left( \frac{x_0}{x} \right)^{\lambda}
\eeq
with $\lambda$ given by \eq{SC2}. The numerical analysis of the value of 
$\lambda$ can be found in Ref. \cite{DUR}. The main conclusion from this analysis is the fact that the value of $\lambda$ is 
sensitive to higher order correction in $\as$. Therefore in this paper we choose to
 fix the value of $\lambda$ from the 
phenomenological approach, see \eq{QSWY}; we consider \eq{SC3} as 
a justification for the use of such a parameterization.

Another observation on the equation for the saturation scale \eq{SC2} is that the value of $\gamma_{cr}$ 
is stable with respect to higher order corrections and almost does not depend on the value of the QCD coupling 
(see Ref. \cite{DUR}).  This fact helps us to solve \eq{SC2} in the case of running $\as$. 
The running of the coupling constant $\as$ leads to an additional dependence on $Q_s$ in the r.h.s. of \eq{SC2}; 
from \eq{SC2} using the explicit form of the running coupling constant we find
\be \label{SC4}
\frac{ d \ln( Q^2_s(W)/\Lambda^2_{QCD})}{d \,\ln(W/W_0)}\,\,=\,\,
{4 \pi \over \beta_2} \, {\chi(\gamma_{cr}) \over 1 - \gamma_{cr}} \, {1 \over \ln(Q^2_s(W)/\Lambda^2_{QCD})} \equiv 
{\delta \over \ln(Q^2_s(W)/\Lambda^2_{QCD})};
\ee
as a result, the dependence on $Q_s(W)$ has become explicit. 
Integrating \eq{SC4} we obtain
\be \label{QSR}
Q^2_s(W)\,\,=\,\,\Lambda^2_{QCD}\,
\,\,\exp\left(  
\sqrt{\,2\,\delta \,\ln(W/W_0)\,\,+\,\,\ln^2(Q^2_s(W_0)/\Lambda^2_{QCD})}\right),
\ee
where $Q^2_s(W_0)$ is the saturation scale at the energy $W_0$ which we used as an initial condition in integrating \eq{SC4}.
Here as well as in the rest of the paper $\Lambda^2_{QCD}$ is defined by  $\as = 4\pi/\beta_2 
\,\ln(Q^2/\Lambda^2_{QCD})$ and in 
numerical applications we took $\Lambda^2_{QCD} = 0.04\ \  {\rm GeV}^2$ with $\beta_2 = 11 - 2/3 \,N_f$ where $N_f = 3$ is the number 
of fermions (number of colors $N_c = 3$). We fix the value of $\delta$ through the empirical value of $\lambda$ as given by \eq{SC3} and 
the value of saturation scale for the $Au$ nucleus at fixed energy of $W= 130$ GeV, $y=0$, corresponding to the cut of $0-6 \%$ of most 
central collisions, $Q_{s0}^2 = 2$ GeV$^2$, so that $\delta = \lambda   \ln(Q^2_{s0}/\Lambda^2_{QCD})$.

The formula \eq{QSR} reproduces all general features expected for the case of running QCD coupling; in particular, one can see 
that the saturation scale (\ref{QSR}) does not depend on the mass number of the nucleus in the limit of high energies \cite{LERYD,MUQA} -- 
the parton wave functions of different nuclei in this limit become universal. 
It is easy to generalize \eq{QSR} to $y \neq 0$ by replacing $\ln(W/W_0)$ by $\ln(W/W_0) + y$; thus we have the following final formula 
 for the case of running $\as$:  
\be \label{QSRUNF}
Q^2_s(y,W) = \Lambda^2_{QCD}\exp\left(\sqrt{2\lambda \ln(Q^2_{s0}/\Lambda^2_{QCD})[\ln(W/W_0) + y] + \ln^2(Q^2_s(W_0)/\Lambda^2_{QCD})
}\right).
\ee

\subsection{Formulae for the multiplicities}

To derive the final expressions for the multiplicity it is convenient to
re-write \eq{MULTI} using the fact that the main contribution to \eq{MULTI} is given by 
two regions of integration over $k_t$: $k_t \,\ll\,p_t$ and $|\vec{p}_t \,-\,\vec{k_t}|\,\ll\,p_t$; this leads to 
\be \label{MUXG}
{dN \over dy} & = & {1 \over S}\ \ \int d p_t^2 \left( E {d \sigma \over d^3 p} \right) 
           \,    = \, {1 \over S}\ \ {4 \pi N_c \as \over N_c^2 - 1}\ \times \nonumber \\
& \times  & \int {d p_t^2 \over p_t^2}\ \left(
\,\varphi_{A_1}(x_1, p_t^2)\,\, \int^{p_t} \, d k_t^2 \
 \varphi_{A_2}(x_2, k_t^2) \,\,+\, \,\varphi_{A_2}(x_2, p_t^2)\,\, \int^{p_t} \, d
k_t^2  \ \varphi_{A_1}(x_1, k_t^2) \, \right) \,\, = \nonumber \\
&=&  {1 \over S}\ \ {4 \pi N_c \as \over N_c^2 - 1}\
\,\int^{\infty}_0\,\,\frac{
d\,p^2_t}{p^4_t}\,\,x_2G_{A_2}(x_2,p^2_t)\,\, x_1G_{A_1}(x_1,p^2_t)\,\,,
\ee
where we integrated by parts and used \eq{XGPHI}.
In the KLMN treatment \cite{KN,KL,KLM,KLN}  we assumed a simplified form of $xG$, namely,
\be \label{XGSAT}
xG(x;p^2_t)\,\,=\,\,\left\{\begin{array}{l}\,\,\,\, {\displaystyle{\kappa \over \as(Q^2_s)}}\,
S\,
p^2_t\,\,(1\, -\, x)^4, \,\,\,\,  \quad \quad p_t\,<\,Q_s(x)\,\,; \\ \\
\,\,\,\, {\displaystyle{\kappa \over \as(Q^2_s)}}\,
S\,Q^2_s(x)\,\,(1\, -\, x)^4, \,\,\,\, \quad \quad p_t\,>\,Q_s(x)\,\,;
\end{array}
\right.
\ee
where the normalization coefficient $\kappa$ has been determined from the RHIC data on gold-gold collisions. 
 We introduce the
factor $(1 \,-\,x)^4$ to describe the fact that the gluon density is small at
$x\,\rightarrow\,1$ as described by the quark counting rules \cite{Brodsky,Matveev}.

We have checked 
that the simplified form of \eq{XGSAT} is adequate for the calculations of multiplicity since it is dominated by the low momenta region. 
At high $p_t$ and small $x$, it was shown  \cite{KLM} that the quantum effects of the anomalous 
dimension could be extremely important. 
However, at moderate values of $x$ the simple form of \eq{XGSAT} was used to calculate the $p_t$ 
spectra in proton-proton and electron-proton collisions in Ref. \cite{szc} and the results appear very 
encouraging.

Having in mind \eq{XGSAT}, let us divide the   $p_t$ integration in
\eq{MULTI} in  three different regions: 
\begin{enumerate}
\item \quad $ p_t\,\,<\,\,Q_{s, min} $

In this region both parton densities for $A_1$ and $A_2$   are in the
saturation region. This region of integration gives  
\beq \label{M1}
\frac{d N}{d y} \,\,\propto\,\,\frac{1}{\as}\,S\,Q^2_{s, min}\,\,\propto
\frac{1}{\as}\,\,N_{part}(A_1)
\eeq
where we have used the fact that the number of participants is proportional
to $S Q_s^2$, where $S$ is the area corresponding to a specific centrality cut.

\item \quad $ Q_{s,max}\,\,>\,\,p_t\,\,>\,\,Q_{s, min} $

For these values of $p_t$ we have saturation regime for the nucleus $A_2$  for all positive  rapidities
 while the nucleus $A_1$  is in the normal DGLAP
evolution
region. Neglecting anomalous dimension of the gluon density below $Q_{s,max}$, 
we have $\varphi_{A_1}(x_1, k_t^2)
\,\,\propto\,\,\frac{1}{\as}\,\,S\,\,Q_{s,min}/k^2_t$ which  for $y > y_c$ leads to 
\beq \label{M2}
\frac{d N}{d y} \,\,\propto\,\,\frac{1}{\as}\,S\,Q^2_{s, min}\,\ln
\frac{Q^2_{s,max}}{Q^2_{s,min}}\,\,\propto
\frac{1}{\as}\,\,N_{part}(A_1)\,\,\ln
\frac{Q^2_{s,max}}{Q^2_{s,min}}
\eeq
This region of integration will give the largest contribution.

\item \quad $ \,p_t\,\,>\,\,Q_{s, max} $

In this region the parton densities in both nuclei are in the DGLAP
evolution region.

\end{enumerate}

\vskip0.3cm

Substituting \eq{XGSAT} into \eq{MUXG} we obtain the following formula \cite{KL}: 
\beq \label{MSTR1} 
\frac{dN}{d
y}\,\,=\,\,Const\, \times \, S\,\,Q^2_{s,min}(W,y)\,\,\frac{1}{\as(Q^2_{min}(W,y))}\,\times
\eeq
$$
\left[ 
\left(1 - \frac{Q_{s,min}(W,y)}{W}\,e^{y}\right)^4 
\,\,+\,\,\{\,\ln(Q^2_{max}(W,y)/Q^2_{s,min}(W,y))\,\,+\,\,1\,\}\,\left(1
-
\frac{Q_{s,max}(W,y)}{W}\,e^{y}\right)^4\,\right].  \nonumber
$$

One can see two qualitative properties of \eq{MSTR1}. For $y\,>\,0$ and close to
the fragmentation region of the nucleus $A_1$ , $Q_{s, min} = Q(A_1)$ and the multiplicity is
proportional to $N_{part} (A_1)$, while in the  
fragmentation region of the nucleus $A_2$($y < 0$) $Q_{s, min} = Q(A_2)$ and $dN/dy \,\propto\, N_{part}(A_2)$. We thus recover some 
of the features of the phenomenological `wounded nucleon' model \cite{WNM}.

 \section{Predictions}

\subsection{Choice of the phenomenological parameters}

As discussed above our main phenomenological parameter is the saturation momentum. An estimate of the value 
of the saturation momentum can be found from the following condition: the probability of interaction in the target (or "the 
packing factor" of the partonic system) is equal to 
unity. The packing factor can be written in the following form:
\beq \label{PFN}
P.F.\,\,=\,\,\frac{8\,\pi^2\,N_c\, \as(Q^2)}{(N^2_c - 1)\,Q^2}\,\,\frac{xG(x,Q^2)}{\pi\,R^2}\,\,=\,\,\sigma\,\,\rho
\eeq
where $\sigma$ is the cross section for dipole - target interaction (the size of the dipole is about 1/Q) and 
$\rho$ is the (two-dimensional) transverse density of partons inside the target of size $R$ (see e.g. Refs. \cite{KN,KL} for details).

In the case of the nucleon we do not know the value of $R$ or, in other words, we do not know the area which is 
occupied by the gluons ($S_N \,=\,\pi R^2$). However, we have enough information to claim that this area is less 
than the area of the nucleon  ( $R$ is less than the electromagnetic radius of the proton). To substantiate this claim, let us recall for example 
the constituent quark model in which the gluons are distributed in the area determined by the small (relative to the size of the nucleon) size of the 
constituent quark.  Having all these uncertainties in mind we use the phenomenological Golec-Biernat and Wuesthoff 
model \cite{GW} to fix the value of the saturation moment in the case of the nucleon target.  Namely, the value of 
the saturation moment for proton is equal $Q_s(P;y=0,W=200\,{\rm GeV}) \,=\,0.37\,{\rm GeV}^2$. In Ref. \cite{KLND} this value of the proton saturation momentum was used to describe the 
deuteron-gold collisions at RHIC energies.

In the case of the nuclear target the saturation momentum can be found from the expression for the packing factor
\beq \label{PFA} 
P.F. 
\,=\,\sigma\,\rho_A\,=\,\sigma\,\rho_N\,\frac{\rho_{part}}{2}\,\frac{S_N}{S_A}\,=\,Q^2_S(N)\,\frac{\rho_{part}}{2}
\,\frac{S_N}{S_A}
\eeq
As we have discussed we do not know the last factor ($S_N/S_A$) and therefore, \eq{PFA} cannot help us to 
determine the value of the saturation momentum for a nucleus. We fixed the value of the saturation momentum from 
the description of the RHIC data on the multiplicity in gold-gold collisions, namely, 
$Q_s^2({\rm Gold},y=0,W=130\,GeV)\,=\,2.02\,{\rm GeV}^2$ for the centrality cut $0 - 6 \%$ (see Refs. \cite{KN,KL} for 
details).

As far as energy dependence of the saturation scale is concerned, we used \eq{QS} and \eq{QSRUNF} with $\lambda$ 
given by the Golec-Biernat and Wuesthoff model ($\lambda = 0.252$).  However, we have to admit that the 
perturbative QCD estimates described above would lead to a larger value of $\lambda$ :  $\lambda \,\approx\,\, 0.37$. Such an uncertainty 
in the value of $\lambda$ leads to an error of about $12 - 15  \%$ in our prediction for the proton-nucleus  
and nucleus-nucleus collisions at the LHC energies. For the proton-proton interaction it could generate an error as big as  
about $50 \%$.  

In our main formula given by \eq{MSTR1} we have to fix the normalization factor. As discussed in Ref. 
\cite{KL} theoretical estimates lead to a value of $Const$ in  \eq{MSTR1} which appears quite close to the value extracted  
by comparison with the RHIC data. In this paper we use the 
same normalization factor $Const$ as in \cite{KL}, namely $Const \,S\,Q^2_{s,min}(W = 
200\,GeV,y=0)\,=\,0.615\,N_{part}$. We also need to note that the experimental measurement are done at fixed pseudo-rapidity $\eta$, 
not rapidity $y$; therefore, as discussed in Ref. \cite{KL} we have to use the relation between $\eta$ and $y$, and to multiply \eq{MSTR1} by 
the Jacobian of this transformation $h(\eta, Q_s)$; see \cite{KL} for details and explicit expressions. 

Proton-proton collisions present an additional problem caused by the deficiency of the geometrical interpretation of $pp$ cross section. 
As mentioned above (see \eq{MULTI}) we calculated 
the ratio of the total inclusive cross section to the geometrical area of the interaction. This ratio is the 
measured multiplicity in the case of hadron-nucleus and nucleus-nucleus collisions. In the case of hadron-hadron 
interaction the multiplicity  is the ratio of the inclusive cross section divided by the inelastic cross section. 
As discussed above, for $pp$ interactions we do not know the relation between the  interaction area and the value of the inelastic 
cross section.  To evaluate  multiplicities in proton-proton interactions we do the following: (i) fix the ratio
$S_N/\sigma_{in}$ at $W = 200\,GeV$ using the data for $dN/dy (y=0)$; and (II) assume 
$S_N/\sigma_{in}\,\,\propto\,1/\sigma_{in}$ as far as the energy dependence is concerned.  In other words we 
assume that the area $S_N$ does not depend on energy.  The energy dependence of the inelastic cross section, including energies outside 
of the region accessible experimentally at present, 
was taken from Ref. \cite{GLMSF}.

\subsection{Proton - proton collisions}

\subsubsection{Rapidity distribution}

\begin{figure}
\begin{center}
\epsfxsize=12.5cm  
\hbox{ \epsffile{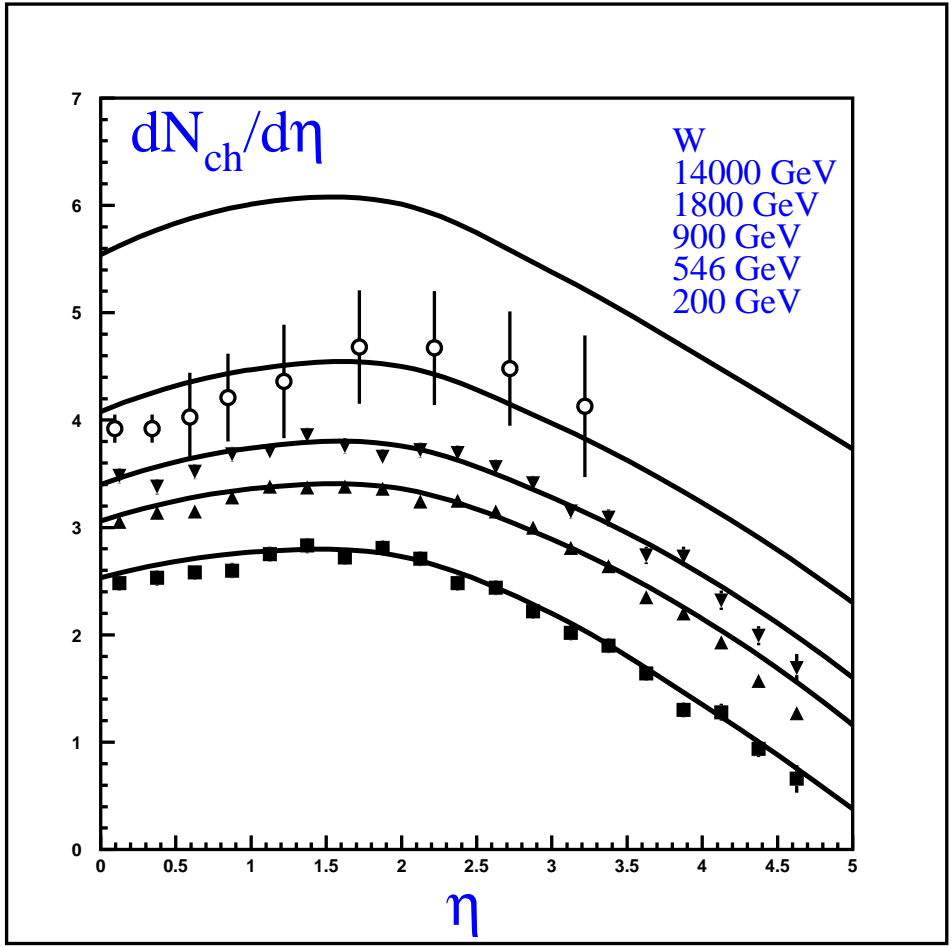}}
\end{center}
\caption{Rapidity dependence $dN/d\eta$ of charged hadron multiplicities in proton -  proton (antiproton) 
collisions  as a function of the pseudorapidity at different energies. The data are taken from Ref. \cite{RTABL}.}
 \label{dndyp}
\end{figure}

\begin{figure}
\begin{center}
\epsfxsize=12.5cm
\hbox{ \epsffile{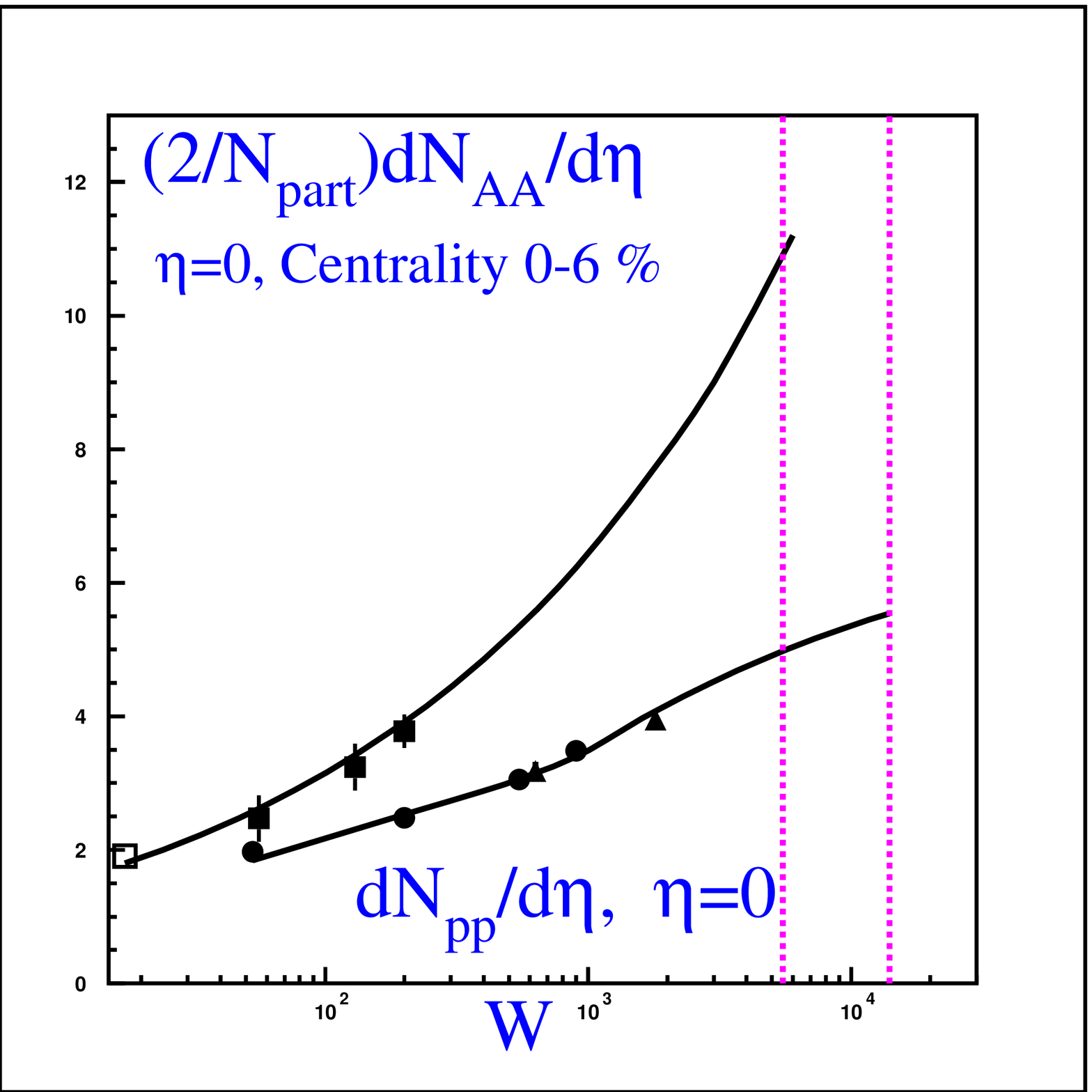}}
\end{center}
\caption{Energy  dependence of charged hadron multiplicity $dN/d\eta$ at $\eta =0$ in proton -  proton 
(antiproton) collisions and of charged hadron multiplicities per participant pair  $(2/N_{part})\,dN/d\eta$ at $\eta =0$ for 
central nucleus-nucleus  
collisions. The vertical dotted lines mark the LHC energies for 
nucleus-nucleus collisions ($W = 5500\,GeV$) and for proton-proton collisions ($W= 14000\,GeV$).
collisions. The experimental data are from Ref. \cite{RTABL,REF1}.}
 \label{dndy0}
\end{figure}

\fig{dndyp} shows the calculated pseudorapidity distributions for the proton - proton (antiproton) collisions. The agreement 
with the experimental data is quite good despite the fact that $pp$ collisions present special difficulties for 
our approach since the value of the saturation momentum is rather small and non-perturbative corrections 
could be essential.  We would like to point out however that the value of the saturation momentum for the proton 
reaches $\approx\,\,1\,GeV$ at the LHC energy. Our experience with RHIC data suggests that at such value of the 
saturation momentum our approach could apply with a reasonable accuracy. 
We thus expect that the very first $pp$ data from the LHC can provide an important test of the CGC ideas.  
In  \fig{dndy0} we plot the value of $dN/d \eta$ at $\eta = 0$ as a function of energy (in this plot we included also the available  
data at lower energies). The agreement with the experiment is seen to be quite good.

\begin{boldmath}
\subsubsection{Total multiplicity}
\end{boldmath}

\begin{figure}
\begin{center}
\epsfxsize=12.5cm
\hbox{ \epsffile{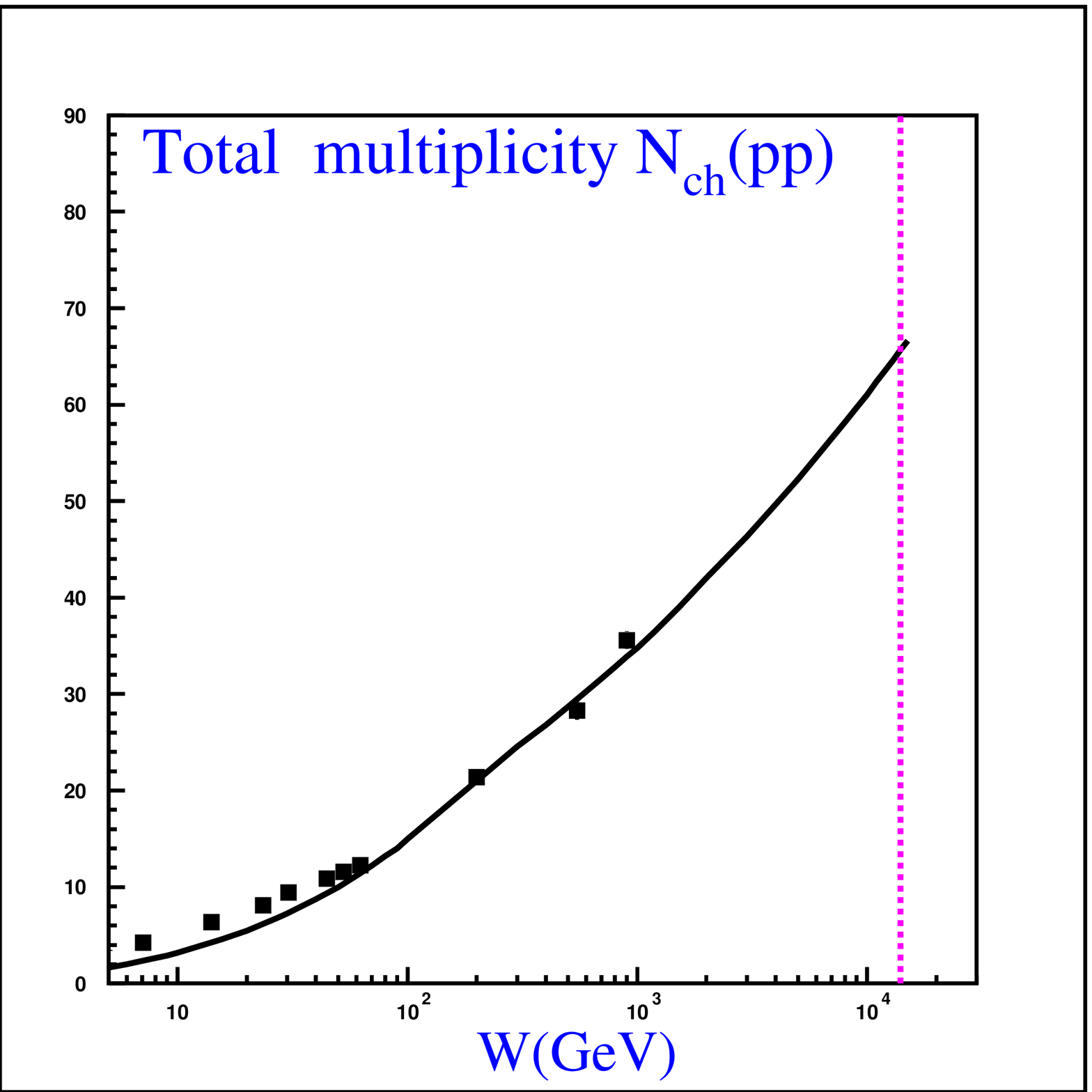}}
\end{center}
\caption{Energy  dependence of total multiplicity   in proton -  proton
(antiproton)  collisions.  The vertical dotted line marks the LHC 
energies  for proton-proton collisions ($W= 14000\,GeV$).
collisions. The experimental data are taken from Ref. \cite{REF2}.}
 \label{N}
\end{figure}

Integrating \eq{MSTR1} over $\eta$ in the entire region of $\eta = - \ln W \,\div + \ln W$ we can  calculate the total multiplicity 
in proton-proton ( antiproton) collisions. In \fig{N} we present our calculation together with the experimental 
data taken from Refs. \cite{REF2}; a good agreement with the data is seen in a wide range of energies. We would 
like to remind however that our predictions for the LHC energies could be as much as 1.5 times larger due to uncertainties in the 
energy behavior of the saturation scale discussed above. 

\subsection{Nucleus-nucleus  collisions.}

\begin{boldmath}
\subsubsection{Rapidity distribution $dN/dy$}
\end{boldmath}

\begin{figure}  
\begin{center}
\epsfxsize=12.5cm
\hbox{ \epsffile{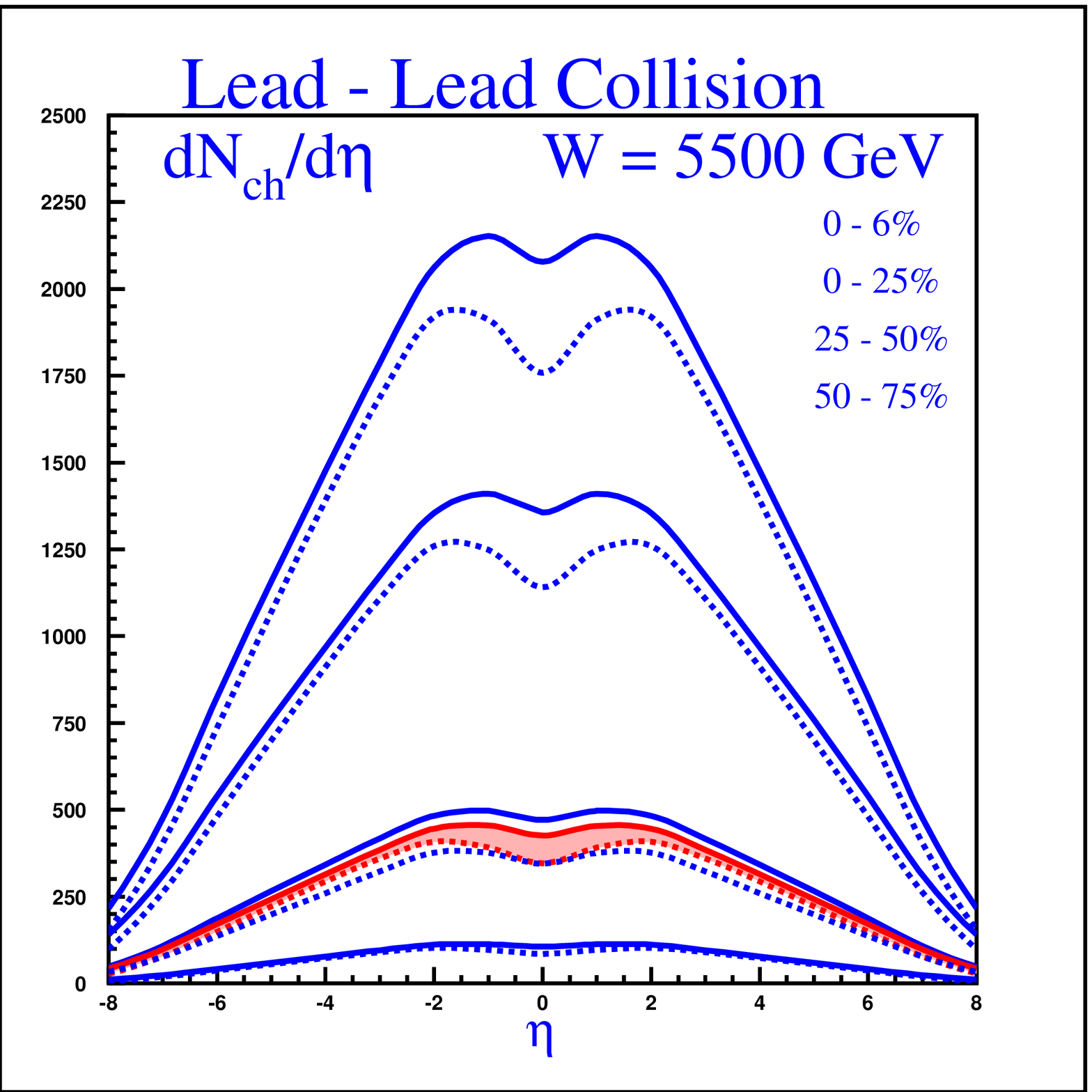}}
\end{center}
\caption{Rapidity   dependence of $d N/d \eta$   lead-lead collisions at the LHC energy at different centrality 
cuts. The solid lines corresponds to the prediction using \eq{QS} for the energy dependence of the saturation 
scale while the dotted lines show the predictions for \eq{QSRUNF} for running QCD coupling. The shadowed area 
shows the prediction for the minimal bias event.}
 \label{dndya}
\end{figure}

Our prediction for lead-lead collision at the LHC energy is plotted in \fig{dndya}. The two sets of curves ( solid 
and dotted) describe the cases of fixed and running QCD coupling respectively. We consider the two predictions as 
the natural bounds for our predictions, and expect the data to be in between of these two curves. However, we would 
like to mention again that our predictions have systematic errors of about $12 \div 15\%$  due to  uncertainties in 
the energy dependence of the saturation scale.

\begin{boldmath}
\subsubsection{Centrality dependence: $(2/N_{part})\,(dN_{ch}/d \eta)$}
\end{boldmath}

\begin{figure}
\begin{center}
\epsfxsize=12.5cm
\hbox{ \epsffile{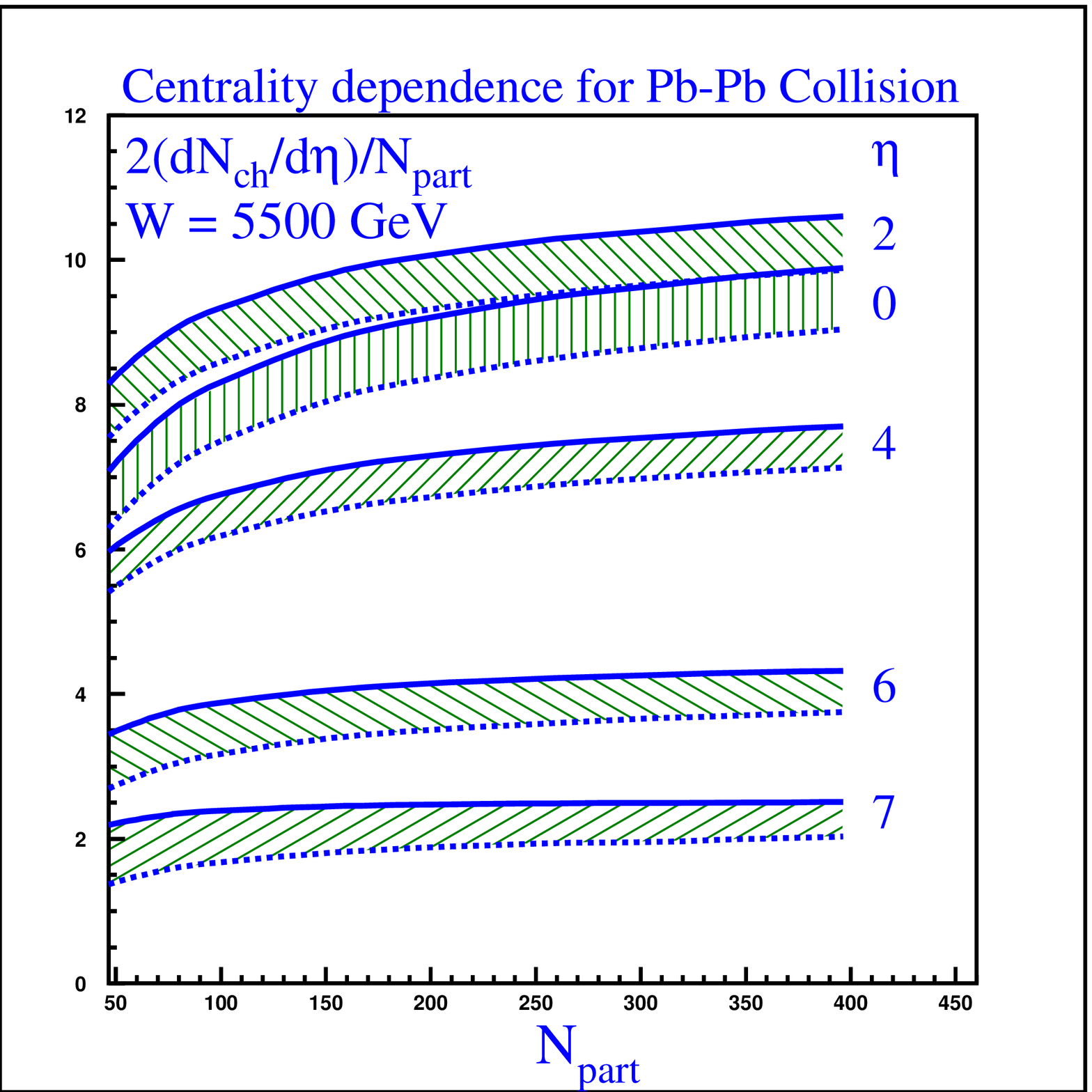}}
\end{center}
\caption{$N_{part}$    dependence of $(2/N_{ch})\,d N/d \eta$   for lead-lead collisions at the LHC energy at 
different rapidity
cuts. The solid lines correspond to the prediction using \eq{QS} for the energy dependence of the saturation   
scale while the dotted lines show the predictions for \eq{QSRUNF} for running QCD coupling. The shadowed areas
show the spread  of our predictions.}
 \label{dndycent}
\end{figure}

\fig{dndycent} shows our predictions for the $N_{part}$ dependence of the $(2/N_{part})\,(dN_{ch}/d \eta)$.
This observable provides the most sensitive test of the value of the saturation scale and its dependence on the density of 
the participants. 

\fig{dndy0} shows the energy dependence of $(2/N_{ch})\,(dN/d \eta)$ at $\eta = 0$.  One can see that we are able 
to describe the current experimental data. Note that if we neglect the difference between rapidity and pseudorapidity, $(2/N_{ch})\,(dN/d \eta)$ at $\eta = 
0$ is given by a very simple formula \cite{KLN}:  
\be \label{DNDY0}
\frac{2}{N_{part}}\,\frac{dN_{ch}}{d y}\,|_{y = 0}\,\,&=&\,\,0.87 \,\left( \frac{W}{W_0} \right)^{\lambda}
\,\ln(Q_s^2(A,W,y=0)/\Lambda_{QCD}^2) = \nonumber\\
 &=& 0.87\,\left( \frac{W}{130} \right)^{0.252}\,\left( 3.93\,\,+\,\,0.252 \,\,\ln(W/130) \,\right)
\ee 
This formula is in good agreement with the existing experimental data.

\subsection{Proton - nucleus  collisions}

\begin{figure}
\begin{center}
\epsfxsize=12.5cm
\hbox{ \epsffile{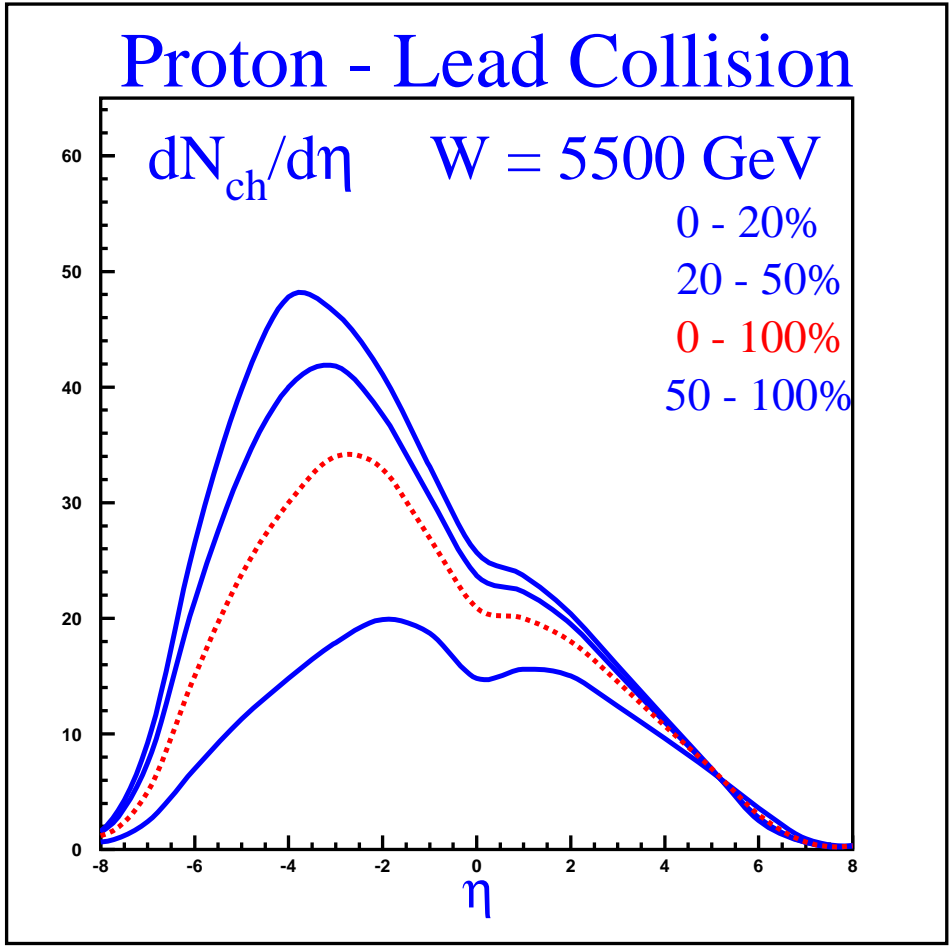}}
\end{center}
\caption{Rapidity   dependence of $d N/d \eta$   proton-lead collisions at the LHC energy at different centrality
cuts. The dotted line corresponds to the minimal bias event. }
 \label{dndypa}
\end{figure}

\fig{dndypa} shows our prediction for the proton-nucleus collisions at W=5500\,GeV.
In section 2 
we described the procedure of computing
the number and density of participants in this case; to evaluate the relevant value of the saturation momentum, we take account 
of the energy dependence to extrapolate from RHIC to the LHC energy.  
For example, the density of participants $\rho_{part} \approx 1.84 \ {\rm fm}^{-2}$ corresponds to  
the saturation scale of $Q_s^2 \approx 2 \,{\rm GeV}^2 (5500/200)^{0.252}$ \,$ \approx 4.6 \,{\rm GeV}^2$.

\section{Conclusions}

\begin{figure}
\begin{center}
\epsfxsize=12.5cm
\hbox{ \epsffile{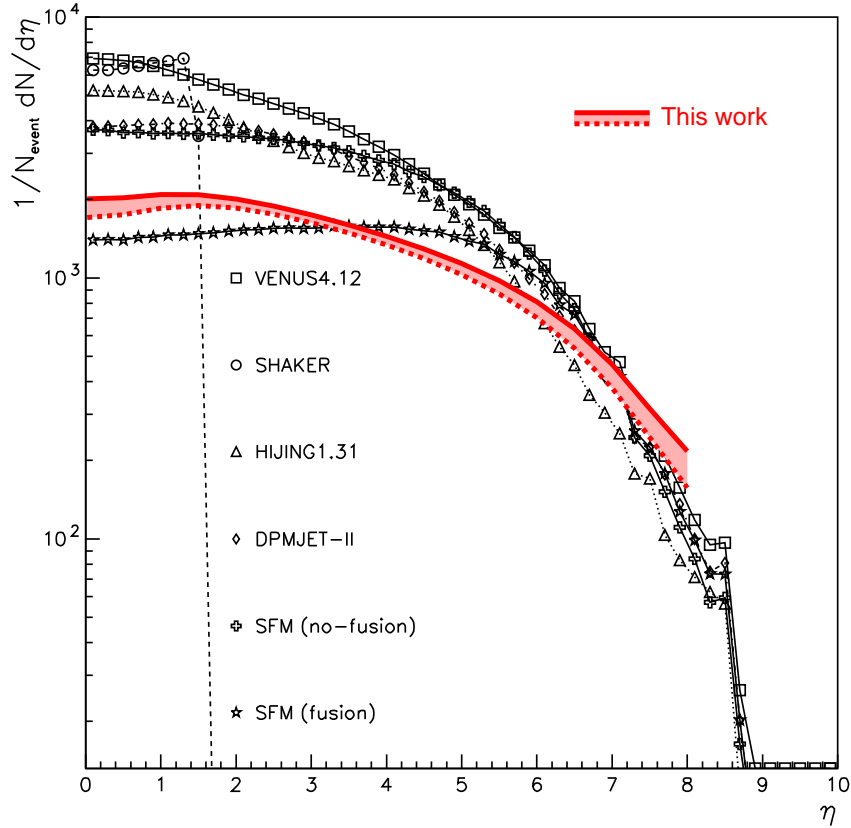}}
\end{center}
\caption{Comparison of our predictions for charged hadron multiplicities in central ($b \leq 3$ fm) $Pb-Pb$ collisions with the results from other approaches, as given in Ref.\cite{REVAP}}
 \label{mult}
\end{figure}

In this paper we have provided a complete set of predictions for the multiplicity distributions at the LHC basing on the CGC approach. 
In our approach, parton saturation results in a relatively weak, compared to most other approaches,   
dependence of the multiplicity on energy. As one can see from \fig{mult} we expect rather small number of produced hadrons in comparison with 
the alternative approaches.  What is the uncertainty in our predictions? We would like to recall the estimates for the uncertainty 
in our calculations at the LHC energies given above: $12 \div 15\%$ for nucleus-nucleus and hadron-nucleus  
collisions, and a large value of $40 \div 50 \%$ for the proton-proton collisions. These uncertainties arise from the 
poor theoretical knowledge of the energy dependence of the saturation scale, and in the case of $pp$ collisions also from the 
uncertainties in the application of the geometrical picture. 

We hope that our estimates will be useful for the interpretation of the first results from LHC experiments.  As illustrated in \fig{mult}, a  
measurement of multiplicity at the LHC will provide a very important test of the CGC approach. 

\section{Acknowledgements}

The work of D.K. was supported by the U.S. Department of Energy under Contract No. DE-AC02-98CH10886. 
E.L. and M.N. are grateful to the RIKEN-BNL Research Center and the Nuclear Theory Group at BNL for hospitality 
and support during the period when this work was done. The work opf E.L. was supported in part by the grant of Israeli Science Foundation 
,founded by Israeli Academy of Science and Humanity.

\end{document}